\documentclass{aa}
\usepackage{psfig}
\usepackage{graphicx}

\def\bron{GRS 1747-312}
\def\ecs{erg~cm$^{-2}$s$^{-1}$}
\def\lum{erg~s$^{-1}$}

\begin{document}

\title{Bursts, eclipses, dips and a refined position for the luminous
low-mass X-ray binary in the globular cluster Terzan 6}

\titlerunning{The luminous LMXB in Terzan 6} 
\authorrunning{J.J.M. in 't Zand, F. Hulleman, C. Markwardt et al.}

\author{J.J.M.~in~'t~Zand\inst{1,2}
\and F.~Hulleman\inst{2}
\and C.B.~Markwardt\inst{3,4}
\and M.~M\'{e}ndez\inst{1}
\and E.~Kuulkers\inst{5}
\and R.~Cornelisse\inst{1,2}
\and J.~Heise\inst{1,2}
\and T.E.~Strohmayer\inst{3}
\and F.~Verbunt\inst{2}
}

% \offprints{J.J.M. in 't Zand, email {\tt jeanz@sron.nl}}

\institute{     SRON National Institute for Space Research, Sorbonnelaan 2,
                NL - 3584 CA Utrecht, the Netherlands 
	 \and
                Astronomical Institute, Utrecht University, P.O. Box 80000,
                NL - 3508 TA Utrecht, the Netherlands
	 \and
                NASA Goddard Space Flight Center, Code 662, Greenbelt,
                MD 20771, U.S.A.
         \and
                Dept. of Astronomy, University of Maryland, College Park,
                MD 20742, U.S.A.
	 \and
                ESA-ESTEC, Science Operations \& Data Systems Division,
                SCI-SDG, Keplerlaan 1, 2201 AZ Noordwijk, the Netherlands
	}

\date{Received, accepted }

\abstract{\bron\ is a bright transient X-ray source in the globular
cluster Terzan~6 with quasi-periodic outbursts approximately every
4.5~months.  We carried out 2--60~keV target-of-opportunity
observations during eight outbursts with the Proportional Counter
Array on the RXTE satellite, for a total exposure time of 301~ks, and
detect the first unambiguous thermonuclear X-ray bursts from this
source. This identifies the compact accretor in this binary as a
neutron star. The neutron star identification implies that twelve out
of thirteen luminous (above 10$^{36}$~\lum) X-ray sources in Galactic
globular clusters harbor neutron stars, with AC211's nature (in M15)
remaining elusive. We observed 24 transitions of eclipses of the X-ray
emitting region by the companion star and are able to improve the
accuracy of the orbital period by a factor of 10$^4$. The period is
$P=0.514980303(7)$~d. We do not detect a period derivative with an
upper limit of $|\dot{P}/P|=3\times10^{-8}$~yr$^{-1}$. Archival
Chandra data were analyzed to further refine the X-ray position, and
the cluster's center of gravity was re-determined from optical data
resulting in a correction amounting to 2 core radii. We find that
\bron\ is $0.2\pm0.2$ core radii from the cluster center. \keywords{accretion,
accretion disks -- globular clusters: individual: Terzan~6 -- X-rays:
binaries -- X-rays: bursts -- X-rays: individual: \bron\ }}

\maketitle 

\section{Introduction}
\label{intro}

\bron\ is a transient X-ray source that was first detected in 1990
with ROSAT (Predehl et al. 1991) and Granat (Pavlinsky et
al. 1994). Within the uncertainty of 1\arcmin\ it was found to
coincide with the globular cluster (GC) Terzan~6. Later this was
confirmed by a more accurate ROSAT localization (Verbunt et al. 1995).
Terzan~6 is a core-collapsed metal-rich GC which is offset from the
Galactic center by 2\fdg6. The core and half-light radii are 3\farcs3
and 26\arcsec\ (Trager et al. 1995), the distance
9.5$^{+3.3}_{-2.5}$~kpc (Barbuy et al. 1997; Kuulkers et al. 2003). It
is heavily reddened with $E(B-V)=2.24$ (Barbuy et al. 1997) which is
among the highest values of GCs harboring luminous ($>10^{36}$~\lum)
X-ray sources. There are thirteen luminous X-ray sources in twelve
globular clusters in the Galaxy. Thus far, eleven of the thirteen
bright GC X-ray sources were proven to harbor neutron stars accreting
matter from a Roche-lobe filling low-mass companion star. The proof
was delivered by the detection of type-I X-ray bursts. Such bursts are
characterized by fast rises (less than a few seconds), slower
exponential-like decays with e-folding decay times between a few
seconds and a few tens of minutes, and blackbody spectra that exhibit
cooling temperatures during the decay (e.g., Lewin et al. 1993;
Strohmayer \& Bildsten 2003). Type-I X-ray bursts are explained as
thermonuclear flashes on the surfaces of neutron stars (Maraschi \&
Cavaliere 1977; Woosley \& Taam 1976). \bron\ and AC211 in M15 (White
\& Angelini 2001) are the remaining cases for which the nature of the
compact objects needs to be resolved. Knowledge about the nature is of
interest for studies about the evolution of GCs (i.e., the
probabilities of tidal capture and escape probabilities of the black
holes versus the lighter neutron stars), particularly now that there
is evidence that GCs in other galaxies do contain black hole X-ray
binaries (e.g., Angelini et al. 2001 and Sarazin et al. 2001).

A previous study of mostly moderately sensitive X-ray data showed that
the outbursts of \bron\ follow each other relatively quickly: In 't
Zand et al. (2000) detected five outbursts between 1996 and 1999 with
recurrence times of about half a year. It was also established that
\bron\ is one of now six known low-mass X-ray binaries that exhibit
complete X-ray eclipses; the others are EXO~0748-676 (Parmar et
al. 1986, Wolff et al. 2002), MXB~1658-298 (Cominsky \& Wood 1984,
Oosterbroek et al. 2001), Her X-1 (Tananbaum et al. 1972),
XTE~J1710-281 (Markwardt et al. 1999) and AX J1745.6-2901 (Maeda
et al. 1996; Kennea \& Skinner 1996). The measured orbital period for
\bron\ is $12.360\pm0.009$~hr, the eclipse duration 43~min.

We initiated a program of dedicated and sensitive
target-of-opportunity (ToO) X-ray observations with RXTE to search for
signatures regarding the nature of the accretor in \bron, in
particular type-I X-ray bursts which can unambiguously be identified
as thermonuclear flashes in a neutron star atmosphere (like for the
other eleven previously mentioned GC cases), and to measure more
eclipses to improve the knowledge about the orbital period. The
present paper concerns the full analysis of all X-ray bursts, the
timing study of the eclipses, and a discussion of the recurrence
behavior of the transient outbursts. Furthermore, we analyzed archival
Chandra data to determine the most accurate X-ray position thus
far. In connection to this we re-determined the cluster's center of
gravity from optical data and find that the X-ray source is within one
core radius from the center.

\begin{figure}[t]
\psfig{figure=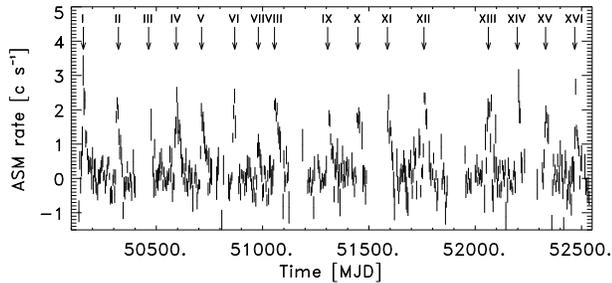,width=\columnwidth,clip=t}
\caption{Time history of \bron\ photon rate in the ASM, between 6
January 1996 and 26 September 2002. A bias level of 0.5~c~s$^{-1}$ was
subtracted. Sixteen outbursts can be recognized. Gaps correspond to
times when the sun crosses the field of view of the source and
observations are impossible. The time resolution is 4~d.
\label{figasmlc}}
\end{figure}

\begin{figure}[t]
\psfig{figure=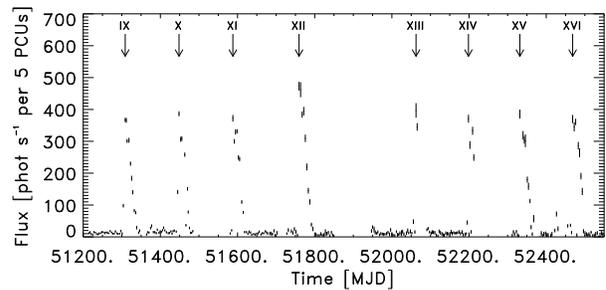,width=\columnwidth,clip=t}
\caption{Time history of \bron\ photon rate during PCA bulge scans
(Swank \& Markwardt 2001; Markwardt \& Swank 2002), between 5 February
1999 and 29 September 2002. Gaps in this plot are also caused by sun
passages.
\label{figbulgelc}}
\end{figure}

\section{ASM and PCA long-term light curves}
\label{lig}

The All-Sky Monitor (ASM) on RXTE monitors \bron\ for about ten months
every year, for up to a few tens of times per day during 90~sec
dwells. The sensitivity is typically a few tens of mCrabs\footnote{1
Crab unit is equivalent to $2\times10^{-8}$~\ecs\ (2--10 keV), for a
power-law spectrum with photon index 2.1} per day of observations
(Levine et al. 1996).  Although limited, the sensitivity is sufficient
to pick up outbursts by \bron\ which typically peak at 30~mCrab. The
resulting light curve is presented in Fig.~\ref{figasmlc}. Sixteen
outbursts are readily recognized by eye, a few (maybe two) were
probably missed at times when the sun passed by \bron\ and
observations were impossible.

Since February 1999 the Proportional Counter Array (PCA, see next
section) is employed in slow semi-weekly scans of a
$16\degr\times16\degr$ field centered on the Galactic center (Swank \&
Markwardt 2001). Fluxes of all point sources in this field are
monitored, including that of \bron. The observation frequency is one
to two orders of magnitude less than that of the ASM observations, but
this is compensated by a considerable gain in sensitivity.
Fig.~\ref{figbulgelc} shows the resulting light curve.

\begin{figure}[t]
\psfig{figure=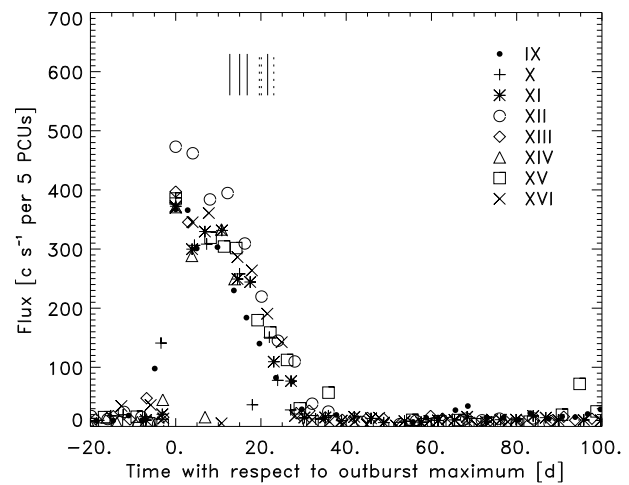,width=1.0\columnwidth,clip=t}
\caption{Same as Fig.~\ref{figbulgelc}, except that the times of the 8
maxima have been subtracted (different symbols for different
outbursts). The vertical lines indicate the times of detections of
type-I X-ray bursts with the PCA (solid) and of burst-like features
with the WFCs (dashed). The three low points during the outburst phase
have been verified to be caused by eclipses.
\label{figbulgefold}}
\end{figure}

If we assume that two outbursts were missed in the ASM data (between
outbursts VIII and IX and between XII and XIII), the average
recurrence time is 136 days. However, the recurrence time varies
between 76 days (between outbursts VII and VIII, see
Fig.~\ref{figasmlc}) and 171 days (between outbursts XI and
XII). There is a clustering of recurrence times in the 130-142 day
range: 7 out of 13 securely determined cases are in this range.

Nearly all outbursts have similar shapes and peak intensities of about
380 c~s$^{-1}$ per 5 PCUs or 2 ASM~c~s$^{-1}$, corresponding to 0.03
Crab units; only two outburst are significantly different: the peak
intensity of outburst VII is half the usual value and outburst XII is
25\% larger. These deviations are accompanied with extremes in the
recurrence time.  Figure~\ref{figbulgefold} shows the PCA scan light
curve folded with respect to the times of maximum of the seven
outbursts.  All PCA-detected outbursts last 4~weeks except for the
last outburst which had low-level activity preceeding the maximum for
one week.

The average intensity from outburst to outburst may accurately be
determined from those outbursts that were well sampled by the PCA
bulge scans and were accompanied by unambiguous measurements of the
recurrence time since the previous outburst. Five outbursts meet these
criteria: X, XI, XII, XV and XVI (including the deviating outburst
XII). We determined the photon fluence per outburst in a
model-independent manner by 1) subtracting a background level as
determined from 3 points preceeding the peak by at least 7 days; 2)
multiplying each data point with the interval time to the next data
point; and 3) summing all values. The resulting fluences divided by
the recurrence time provides average intensities between outbursts
ranging from $50\pm2$ to $58\pm1$ c~s$^{-1}$ per 5 PCUs. The
fractional rms value over the 5 measurements is 5\%.

\section{RXTE-PCA pointed observations}
\label{pca}

\begin{table}
\caption[]{Summary of pointed RXTE observations. The dividing lines 
indicate different outbursts.\label{tabobs}}
\begin{tabular}{lllcr}
Out- & ObsID          & Dates                 & No. & Exp. \\
burst&                &                       & Obs.& (s)  \\   
\hline\hline
VIII   & 30427-01-01&  1998-09-13             &  1  & 3283 \\
\hline	       
IX & 40419-01-01    &  1999-05-12             &  1  & 3356 \\
\hline	       
X  & 40419-01-02    &  1999-09-27             &  2  & 6592 \\
   & 40419-01-03    &  1999-10-13             &  1  & 3584 \\
   & 40419-01-04    &  1999-10-20             &  1  & 2432 \\
\hline	       
XI & 50034-01-01    &  2000-02-18,20,21,24    &  7  & 26546 \\
   & 50034-01-02    &  2000-02-25,26,27,28    &  9  & 34272  \\
   & 50034-01-03    &  2000-03-11,12          &  2  & 2908 \\
   & 50034-01-04    &  2000-03-15             &  1  & 3518 \\
\hline	       
-- & 50034-01-05    &  2000-06-30             &  7  & 11886 \\
   &                &  2000-07-1,2,3,4,5,6    &     &     \\
   & 50034-01-06    &  2000-07-7,8,9,10,11    &  6  & 7791 \\
   &                &  2000-07-12             &     &  \\
\hline					  
XII& 50034-01-07    &  2000-08-10             &  2  & 13521 \\
   & 50034-01-08    &  2000-08-11,12,15       &  3  & 5717 \\
   & 50034-01-09    &  2000-09-09,11          &  3  & 6000$^\dag$ \\
\hline	       
XIII& 60038-01-01   &  2001-05-31             &  1  & 13860 \\
   & 60038-01-03    &  2001-06-03,06          &  2  & 5609 \\
   & 60038-01-04    &  2001-06-10,12,13,16    &  4  & 20942 \\
   & 60038-01-05    &  2001-06-17             &  2  & 9300 \\
   & 60038-01-06    &  2001-07-1,2,5          &  3  & 6556 \\
\hline	       				  
XIV& 60038-01-02    &  2001-10-21,23,24       &  8  &  43983 \\
   & 60038-01-07    &  2001-10-26,27,29,30    &  8  & 48682 \\
   &                &  2001-10-31             &     &  \\
   & 60038-01-08    &  2001-11-02,6,8         &  3  & 14307 \\
   & 60038-01-09    &  2001-11-09             &  1  & 6182 \\
\hline	       				  
XV & 60038-01-10    &  2002-03-14             &  1  & 1377 \\
\hline	       				  
XVI& 60038-01-11    &  2002-07-16             &  1  & 1101 \\
\hline\hline
\normalsize
\end{tabular}
$^\dag$ These observations had bad attitude data and were disregarded.
\end{table}

The Proportional Counter Array (PCA; for a detailed description, see
Jahoda et al. 1996) on RXTE consists of an array of 5 co-aligned
Proportional Counter Units (PCUs) that are sensitive to photons of
energy 2 to 60 keV with a total collecting area of 6500~cm$^2$. The
spectral resolution is 18\% full-width at half maximum (FWHM) at 6 keV
and the field of view is 1$^{\rm o}$ FWHM. During any observation the
number of active PCUs may vary between 1 and 5.

The data on \bron\ were obtained in 80 ToO observations that cover
eight outbursts in 1998-2002, see Table~\ref{tabobs}. The total
exposure time is 301~ks. Thirteen observations (OBSIds 50034-01-05
through 50034-01-06) were scheduled to search for an anticipated
outburst, because the usual trigger observations by the ASM and PCA
bulge scans were absent due to small angular distances to the
Sun. Unfortunately, that outburst did not materialize. On four
outbursts wide coverage was obtained (Feb 2000, Jul 2000, June 2001
and October 2001). Coverage of the other four outbursts was limited to
one or two observations of eclipses.  Many of the individual exposures
were scheduled to cover either an eclipse ingress or egress and not
the full eclipse. An eclipse is rather long and spending much of the
exposure budget on that would have diminished the chances to detect an
X-ray burst.

\begin{figure*}[t]
\includegraphics[width=0.5\columnwidth]{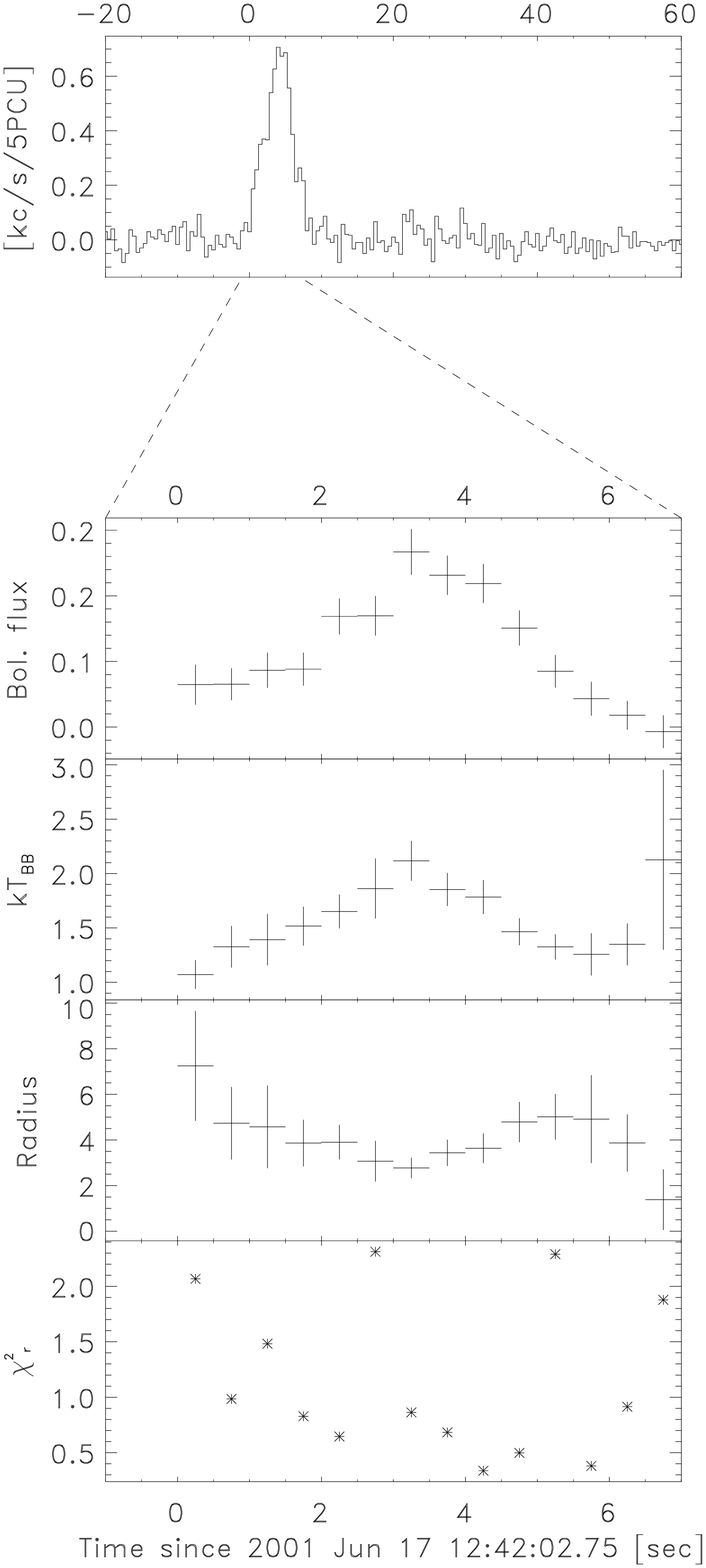}
\includegraphics[width=0.5\columnwidth]{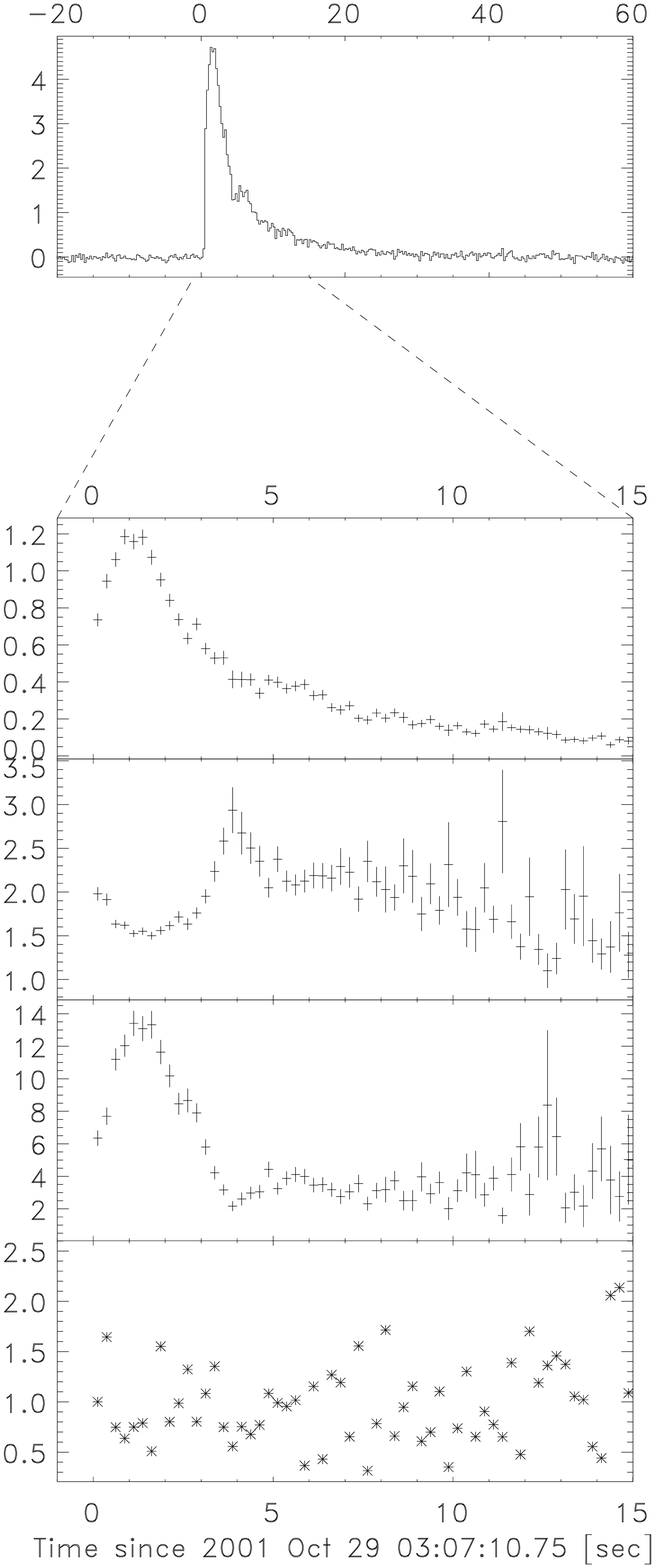}
\includegraphics[width=0.5\columnwidth]{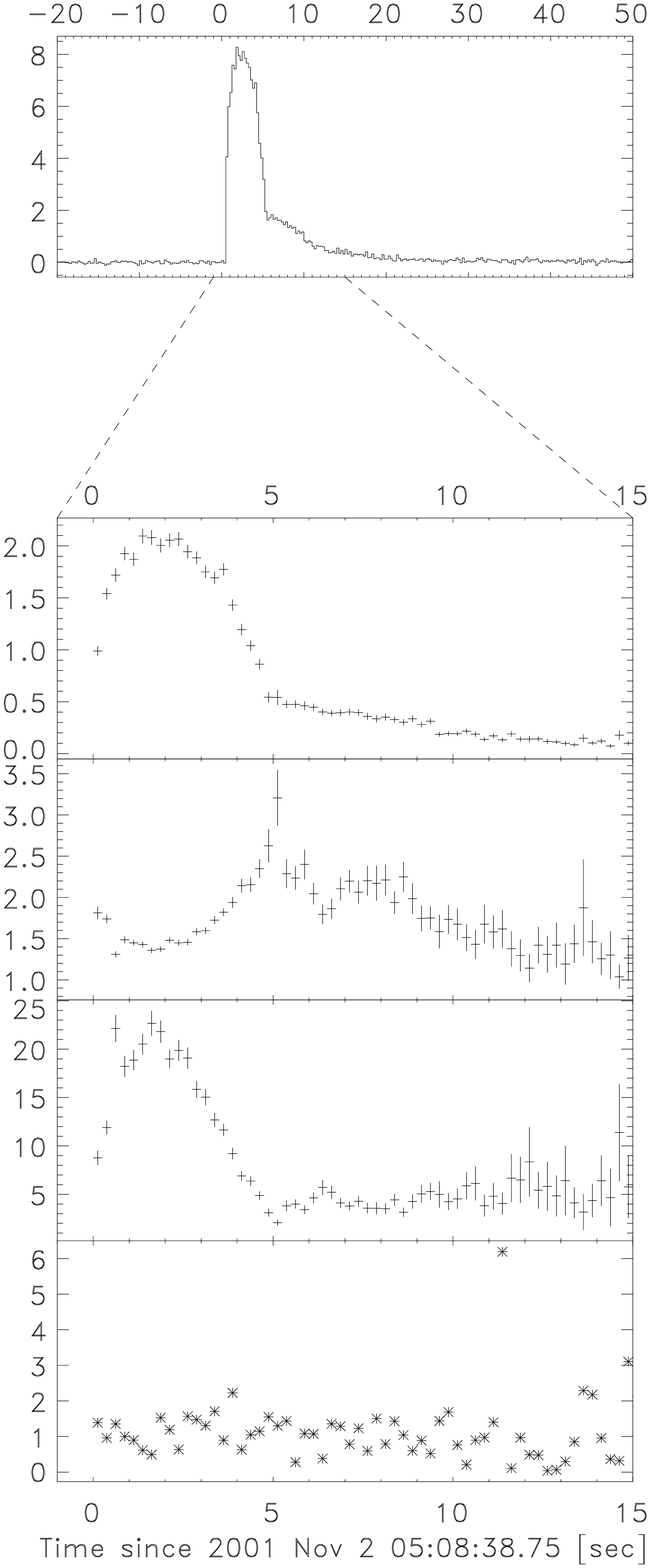}
\includegraphics[width=0.5\columnwidth]{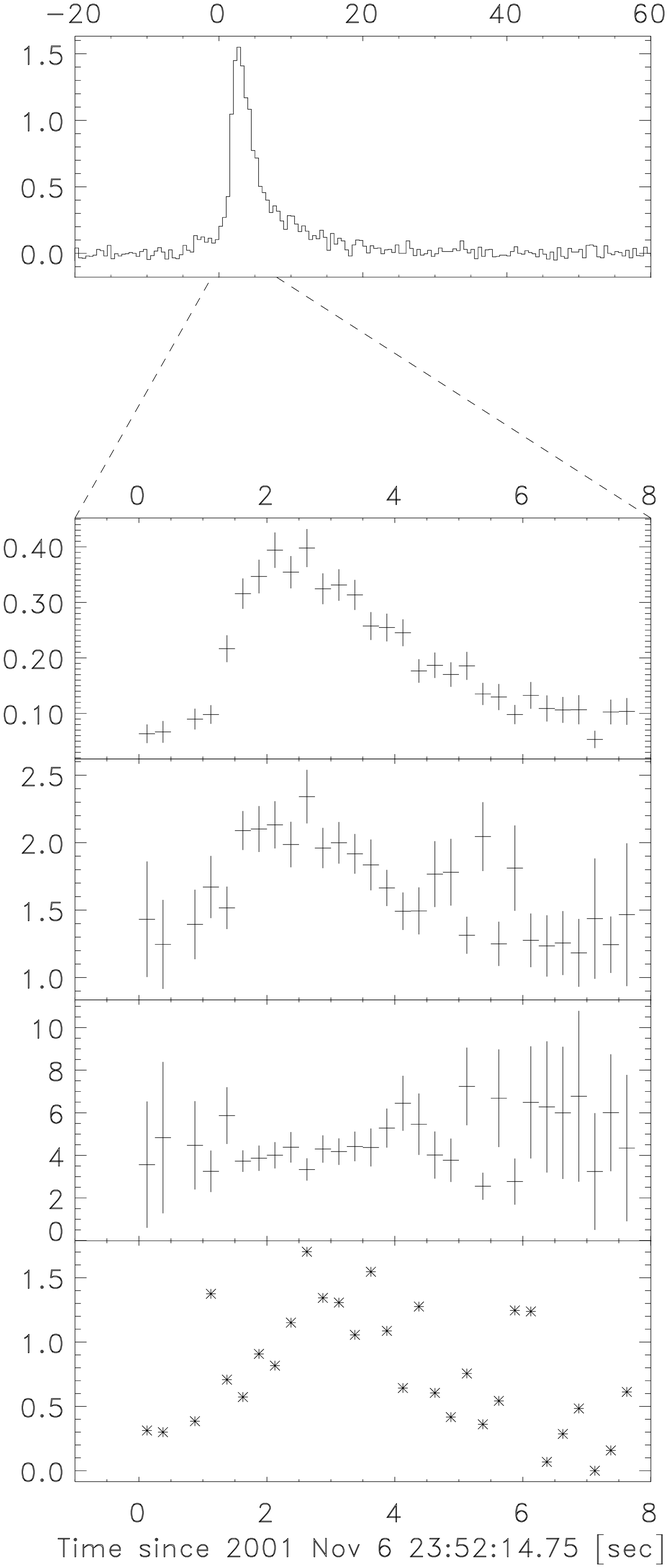}
\caption{Time-resolved spectroscopy of the bursts.  The
upper panel of each burst shows the raw full-bandpass
background-subtracted (including the persistent emission from \bron)
photon count rate, the 2nd panel from above the bolometric
absorption-corrected flux 
in units of 10$^{-8}$~erg~s$^{-1}$cm$^{-2}$,  % new
the 3rd panel the blackbody temperature
in units of keV, % new
the 4th panel the radius of the spherical blackbody emission area
in km % new
(for a distance of 9.5~kpc), and the bottom panel the reduced $\chi^2$
values for the spectral fit to each time-resolved spectrum. We note
that the orbital phases with respect to mid-eclipse for these bursts
are 0.86, 0.28, 0.21 and 0.49 (see Sect.~\ref{ecl}). The abscissa
reference times are in UTC.
\label{fig4bursts}}
\end{figure*}

\section{X-ray bursts}
\label{bursts}

\subsection{Detections}

In outbursts XIII and XIV we were succesful in catching bursts. They
were detected on June 17 (burst no. 1), October 29 (no. 2), November 2
(no. 3) and 6, 2001 (no. 4); all occurred during the decay phase of an
outburst (see Fig.~\ref{figbulgefold}). While the bursts all have
durations of 10 to 15 sec, they rather vary in profile and peak flux
(see top panels of Fig.~\ref{fig4bursts}) and there is a slight
concern that other sources in the field of view are
responsible. Therefore, we applied a localization procedure to those
bursts that employs the slight offsets that exist between PCUs
(Strohmayer et al. 1997). The result is consistent, at the 90\%
confidence level, with all bursts coming from within 0\fdg5 of the
position of \bron. No other bursters are known within this region,
although SAX J1752.3-3138 (Cocchi et al. 2001) is at the
edge. Furthermore, no otherwise (transiently) bright X-ray sources are
known in this region. Given that the bursts all occurred when \bron\
was in an active state, we conclude that they are from the same
source.

Between mid 1996 and mid 2002 the Wide Field Cameras (WFCs; Jager et
al. 1997) on BeppoSAX (Boella et al. 1997) were used to monitor a
$40\degr\times40\degr$ field at 5\arcmin\ resolution around the
galactic center during Feb-Apr and Aug-Oct of each year, with exposure
times of 100~ksec to 1~Msec and observation frequencies of 4 to 67 per
visibility window (e.g., In~'t~Zand 2001). The total net exposure time
is 7~Msec. The bandpass of the WFC is 2 to 28 keV, and the typical
sensitivity is 10~mCrab in one observation. Part of these data were
the basis of the detection of the 12.4~hr orbital period of \bron\
(In~'t~Zand et al. 2000).  The WFC observations revealed the detection
of three burst-like features, with durations of less than 10~s and
peak fluxes comparable to those of the brightest PCA-detected
bursts. The bursts occurred during the decays of outbursts XI and XII
(see Figs.~\ref{figbulgelc} and \ref{figbulgefold}); two were within
half a day from each other. Although the data are of insufficient
statistical quality to allow a detailed spectral analysis, they do
confirm with high localization precision the bursting nature of \bron.

\subsection{Time-resolved spectroscopy}

We carried out a time-resolved spectral analysis of the four
PCA-detected bursts. Preliminary results were presented in Kuulkers et
al. (2003). Event mode data with a resolution of 2$^{-11}$~sec and 64
energy channels were binned to 0.25 through 4~sec, the background was
subtracted as well as the pre-burst persistent emission from \bron, a
correction for deadtime was applied, and the spectral data were,
between 2 and 20~keV, fitted with a single-temperature blackbody
model and constant absorption ($N_{\rm H}=1.4\times10^{22}$~cm$^{-2}$,
see In 't Zand et al. 2000). Furthermore, spectral channels were
combined so that each spectral bin contained at least 15 photons, to
ensure Gaussian statistics and the applicability of the $\chi^2$
statistic. The smallest number of spectral bins was 3.  The results
are summarized in Fig.~\ref{fig4bursts}, together with the raw photon
count rate profiles.  The spectral fits are generally acceptable.  All
bursts show a temperature profile that decreases during the decaying
phase. The blackbody spectrum and the cooling during decay uniquely
identify the bursts as type-I X-ray bursts which are explained as due
to thermonuclear flashes in the upper layers of neutron stars (see
\S1).

Two of the bursts show a time profile for the emission area which is
not constant and anti-correlated with temperature variations. This is
indicative of photospheric radius expansion (by a factor of 3 for
burst 2 and 6 for burst 3) due to near-Eddington fluxes during these
events.  Remarkably, the unabsorbed bolometric peak flux during these
two bursts differs by a factor of 1.8.

\subsection{Search for burst oscillations}

\begin{figure*}[t]
\psfig{figure=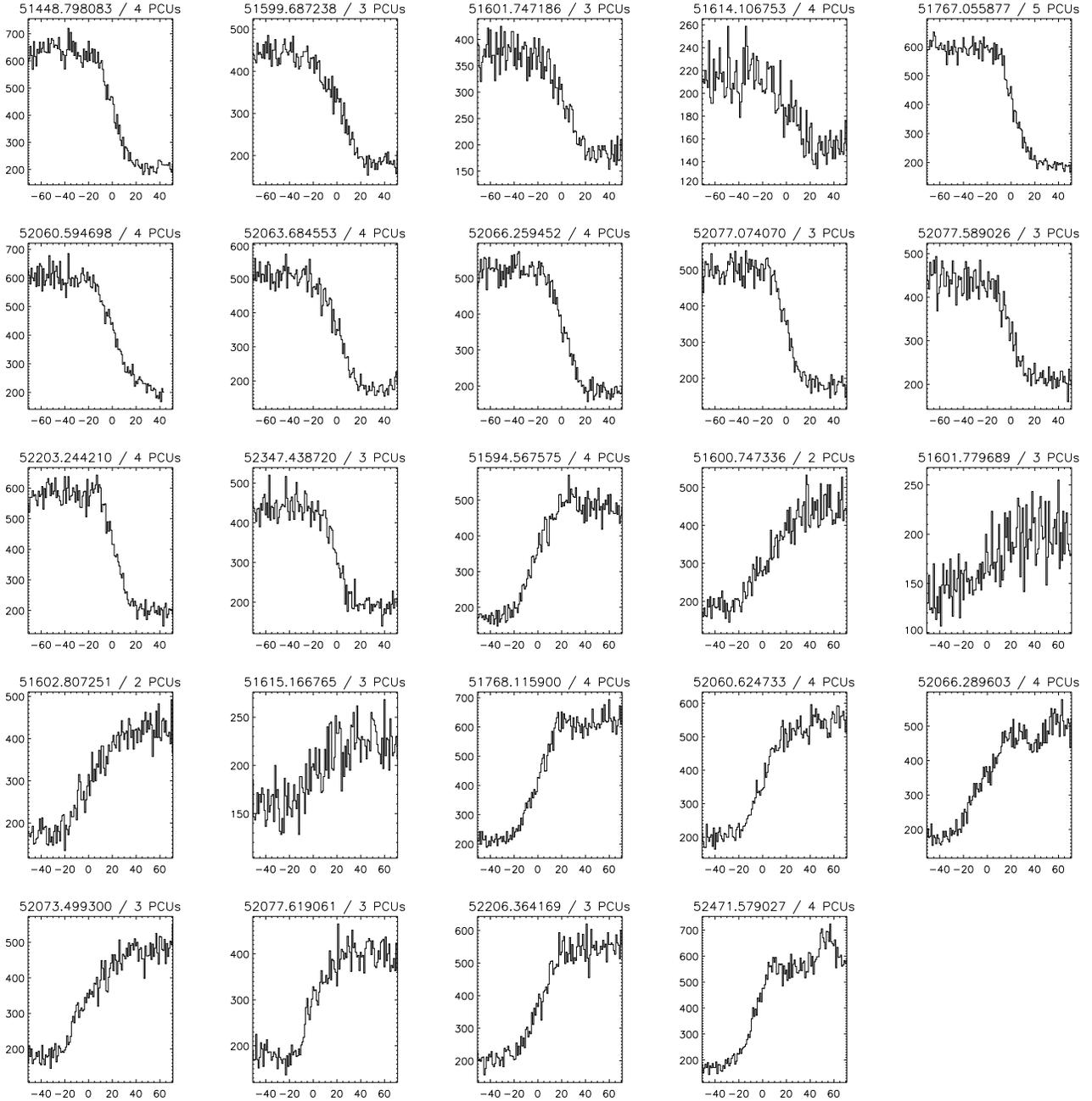,width=2.0\columnwidth,clip=t}
\caption{Light curves of 120~s data stretches around 24 eclipse
transitions in \bron. The abscissa refers to time in second, with a
reference time that is indicated above each plot in MJD (using
Terrestrial Time and corrected to solar system barycenter; TDB), and
the ordinate shows the photon rate normalized for 5 PCUs (note that
the actual number of active PCUs is not constant over all 24 data
stretches and is indicated at the top of each plot).
\label{figlce}}
\end{figure*}

We searched the four X-ray bursts for oscillations. Light curves were
produced at the full time resolution of the data, $2^{-11}$ s,
starting $\sim 5$ s before the onset of the bursts and extending up to
times when the intensity had decayed back to within $\sim 10$\,\% of
the preburst value. We used data for the whole energy band, nominally
$2 - 60$ keV, and from all the detectors that were on during each
burst (three detectors during burst 1, 3, and 4; and four detectors
during burst 2). We calculated one Fourier power spectrum of the whole
burst for each burst, and power spectra of 1, 2, and 4-s intervals
covering the whole duration of the bursts. In the case of the 1, 2,
and 4-s power spectra, to increase the sensitivity to transient
oscillations lasting less than the time length of the power spectra,
we set the beginning of the interval for which we calculated the power
spectrum 0.125, 0.250, and 0.5 s, respectively, after the beginning of
the previous interval (the power spectra were therefore not
independent). For the brightest burst (no. 3), we also calculated
power spectra of 1) 0.25 and 0.5~s intervals, shifting by 1/32 and 1/16~s
at a time; and 2) for $>6$~keV photons only.

We did not find any significant oscillation in the $0.025 - 1024$ Hz
frequency range in any of the bursts. Because of the rapid change in
intensity during the bursts, we calculated upper limits separately for
intervals during which the source intensity was larger than 90\,\% of
the peak intensity, and the rest of the data. For the faintest burst
(no. 1) the observed count rate was a factor of $\sim 7$ lower than
for the brightest burst (no. 3), and therefore the upper limits vary
considerably from burst to burst. The 95\,\% confidence upper limits
for oscillations (for the method, see Vaughan et al. 1994) in
bursts 1 and 4 during the bright and weak parts are 29.8\,\% rms and
63.8\,\% rms, respectively. For bursts 2 and 3, the 95\,\% confidence
upper limits in the bright and weak parts are 11.5\,\% rms and
47.3\,\% rms, respectively.

In the 11 sources that currently are known to exhibit burst
oscillations, the oscillations may reach high amplitudes ($>$50\% rms)
during short intervals (typically 0.1~s) if they occur during the
rising phase and low amplitudes ($<$15\%) during the decaying phase
(for a recent review, see Strohmayer \& Bildsten 2003). Therefore, our
upper limits are not very constraining.

\section{Eclipses and dips}
\label{ecl}

\begin{figure}[t]
\psfig{figure=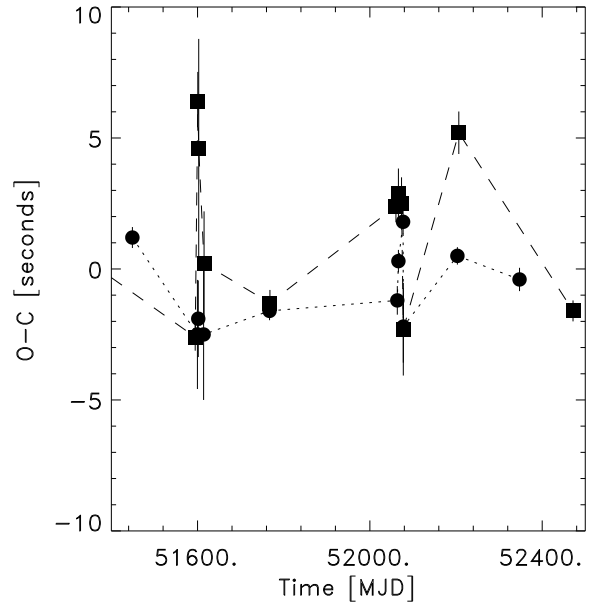,width=1.0\columnwidth,clip=t}
\caption{Observed minus calculated ('O-C') diagram for the eclipse
ingresses (circular symbols connected by a dotted line) and egresses
(rectangular symbols connected by a dashed line) as observed with
RXTE, with respect to the respective ephemerises (see text).
\label{figominc}}
\end{figure}

\begin{figure}[t]
\psfig{figure=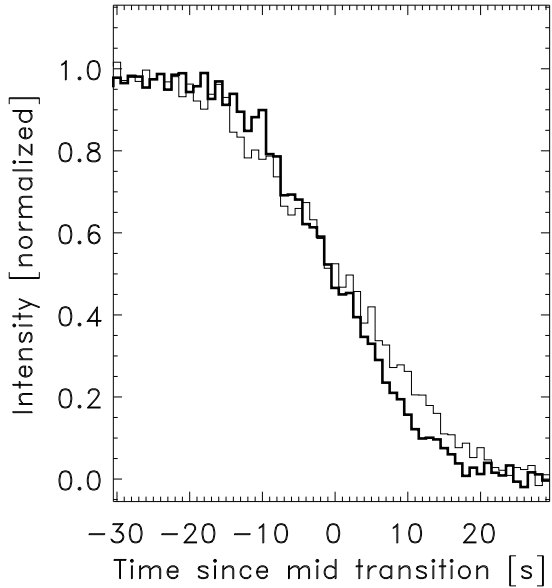,width=1.0\columnwidth,clip=t}
\caption{Average ingress (thick) and egress profile (thin) after
re-normalization of both and mirroring the time axis of the egress
profile. The average profiles have been constructed by aligning
individual profiles at the time of mid transition (see
Table~\ref{tabtiming}).
\label{figfold}}
\end{figure}

\begin{table}[t]
\caption[]{Times and durations of all eclipse transitions
observed thus far with sufficient sensitivity\label{tabtiming}}
\begin{tabular}{llrrr}
\hline\hline
Time             & error & $n^\dag$ & deviation$^\ddag$ & duration$^\ast$ \\
MJD (TDB)        & (s)   &         & (s) & (s) \\
\hline
\multicolumn{5}{c}{ingresses} \\
\hline
51448.798094 & 0.40 & -1199 & +1.2 (+3.0$\sigma$) & $27.9\pm 1.9$ \\
51599.687282 & 2.18 &  -906 & -2.5 (-1.2)         & $56.7\pm14.8$ \\
51601.747211 & 1.43 &  -902 & -1.9 (-1.3)         & $34.1\pm 6.9$ \\
51614.106731 & 2.61 &  -878 & -2.5 (-1.0)         & $22.7\pm 6.8$ \\
51767.055894 & 0.36 &  -581 & -1.6 (-4.5)         & $30.8\pm 2.1$ \\  
52063.684558 & 0.50 &    -5 & -1.2 (-2.2)         & $33.6\pm 3.4$ \\
52066.259477 & 0.47 &     0 & +0.3 (+0.7)         & $29.7\pm 2.6$ \\
52077.074080 & 0.45 &    21 & +1.8 (+3.9)         & $25.5\pm 2.3$ \\
52077.589014 & 1.42 &    22 & -2.2 (-1.6)         & $31.3\pm 5.2$ \\
52203.244242 & 0.36 &   266 & +0.5 (+1.5)         & $25.9\pm 1.8$ \\
52347.438718 & 0.47 &   546 & -0.4 (-0.9)         & $21.6\pm 2.0$ \\
\hline				    		     		 
\multicolumn{5}{c}{egresses} \\	    		     		 
\hline				    		     	 	 
51062.592982$^\times$  & 3    & -1949 & +3.6 (+1.2$\sigma$)& \\ 
51594.567535 & 0.52 &  -916 & -2.6 (-5.0)         & $31.6\pm 3.2$ \\
51600.747403 & 1.14 &  -904 & +6.4 (+5.7)         & $39.3\pm 8.3$ \\
51602.807303 & 4.08 &  -900 & +4.6 (+1.1)         & $68.1\pm22.1$ \\
51615.166779 & 4.68 &  -876 & +0.2 (+0.1)         & $50.5\pm10.0$ \\
51768.115903 & 0.50 &  -579 & -1.3 (-2.6)         & $36.1\pm 3.6$ \\
52060.624742 & 0.60 &   -11 & +2.4 (+3.9)         & $36.2\pm 4.1$ \\
52066.289531 & 0.95 &     0 & +2.9 (+3.1)         & $46.9\pm 6.5$ \\
52073.499250 & 0.97 &    14 & +2.5 (+2.5)         & $41.5\pm 5.9$ \\
52077.619037 & 1.81 &    22 & -2.3 (-1.3)         & $44.7\pm10.4$ \\
52206.364193 & 0.81 &   272 & +5.2 (+6.4)         & $42.2\pm 5.9$ \\
52471.578956 & 0.40 &   787 & -1.6 (-4.1)         & $28.1\pm 2.1$ \\
\hline\hline
\end{tabular}

$^\dag$Number of orbits; $^\ddag$ Deviation with respect to linear fit;
$^\ast$see text for definition; $^\times$BeppoSAX-NFI observation
\end{table}

Twelve eclipse ingresses and twelve egresses were observed, in
twenty-one eclipses distributed over seven outbursts. Two eclipses
were observed completely; for another one the ingress and egress was
covered with a data gap during the eclipse; for the rest only the
ingress or egress was covered. Timing of the transitions provides the
means to improve the accuracy of the orbital ephemeris
significantly. The orbital period thus far had only a 31~sec accuracy
(In~'t~Zand et al. 2000). More accuracy enables studies of changes in
the orbital period. To increase the time baseline, we include the
timing of the egress observed with the BeppoSAX Narrow Field
Instruments on 1998 September 6 (In~'t~Zand et al. 2000). Thus, the
time baseline is 3.9~yr.

Ingress and egress last between 30 and 40 sec. We took the mid points
between minimum and maximum flux as a reference for the timing of the
eclipse transitions. The transition profiles appear too variable to
trust methods that involve profile fitting.  Therefore, we resorted to
the following procedure. As a baseline, we used standard 1 data which
provides 0.125~s resolution. The times were corrected to the solar
system barycenter\footnote{using the {\tt faxbary} routine in {\tt
ftools} version 5.2}. We identified the eclipses in a light curve of
the whole data set at a resolution of 1~s and selected data stretches
of 200~s centered on the initial midpoint determinations (by eye) of
ingress or egress. The light curves of the center 120~s parts of these
data stretches are displayed in Fig.~\ref{figlce}. One ingress had to
be excluded from the analysis (at MJD 52060) because the data stream
stopped before total eclipse was reached; one egress was excluded
because the source was too faint (at MJD 51601). We determined the
weighted average of the initial and final 50~s of each data stretch,
except for the very last egress in which a small flare-like feature
needed to be avoided (see~Fig.~\ref{figlce}). This excludes the
eclipse transitions. The time when the flux reaches the halfway point
between these two averages is determined by identifying the data point
closest to this halfway flux, selecting the 17 data points nearest to
this point, making a least-squares linear fit to these 17 data points,
and finally calculating the interpolated time of the halfway
point. This procedure was developed more or less empirically. It
avoids complications due to changing transition profiles, particularly
near the start and end of each transition. There may be a bias
introduced by the variability of the source outside the eclipse but we
estimate this to be negligible compared to the statistical
accuracy. The statistical errors are of order 1~s (1$\sigma$) in this
analysis. The results of the timing are given in
Table~\ref{tabtiming}. In addition to the timing of the mid-transition
points, we determined a measure for the transition duration by
calculating the time it takes the linear fit to bridge the weight
averages of the initial and final 50~s stretches.

From these data we determine the ephemeris for the ingresses and
egresses. The deviations from these solutions (the ``$O-C$'' values)
are presented in Table~~\ref{tabtiming} and Fig.~\ref{figominc}. Ten
out of the twenty values are larger (in a positive or negative sense)
than 3$\sigma$, but there are no clear trends in $O-C$ and we conclude
that there is no evidence for systematic changes in the orbital
period. Rather we suspect that the larger-than-statistical errors are
due to imperfections of the method. Assuming this is the case, we
multiply the errors of the linear solutions of the ingresses and
egresses by the square-root of the $\chi^2_{\rm red}$ values
($\sqrt{6.6}$ for the ingresses and $\sqrt{15.7}$ for the egresses),
thereby effectively forcing $\chi^2_{\rm red}$ to 1. The ephemeris for
the ingresses is $T_{\rm i}(n) = {\rm MJD~(TDB)}~52066.259473(5) +
0.514980311(8)n$ and for the egresses $T_{\rm e}(n) = {\rm
MJD~(TDB)}~52066.289497(10) + 0.514980275(15)n$. The difference in the
two derived orbital periods is 3.1~ms with an error of 1.7~ms. We take
for the orbital period the weighted mean of 0.514980303(7)~d. This is
only 0.15~s shorter than the previous determination (In~'t~Zand et
al. 2000). We determine an upper limit to the period derivative of
$|\dot{P}/P|=1\times10^{-16}$~s$^{-1}$ or
$3\times10^{-8}$~yr$^{-1}$. The average eclipse duration is 2596(2) s
(or 43.27~min, or 0.05834 orbital periods).

It may be that the larger-than-statistical errors in the
mid-transition times are due to intrinsic properties rather than
imperfections of the method. A similar phenomenon has been observed in
EXO~0748--676.  In addition to a long term period evolution trend, a
significant amount of jitter is detected in that source (Wolff et al.
2002).  Wolff et al. demonstrate that the eclipse phasing of EXO
0748--676 is compatible with a random walk process.  They speculate
that the companion star may become significantly tidally distorted,
enough so that the times of eclipse ingress and egress can be shifted
from their nominal value.  The long term period evolution may be
driven by spin-orbit coupling via the quadrupole moment of the
companion star. In the present analysis we are only interested in the
long-term orbital period evolution of \bron\ and we regard the
jittering as noise.

The average transition profiles are given in
Fig.~\ref{figfold}. Surprisingly, there is a difference between the
average ingress and egress. In particular, the average egress appears
to progress about 1.3 times slower than the ingress. This is confirmed
by the averages of the durations as given per transition in
Table~\ref{tabtiming}: they are $27.3\pm0.8$~s for ingress and
$33.8\pm1.3$~s for egress.

\begin{figure}[t]
\psfig{figure=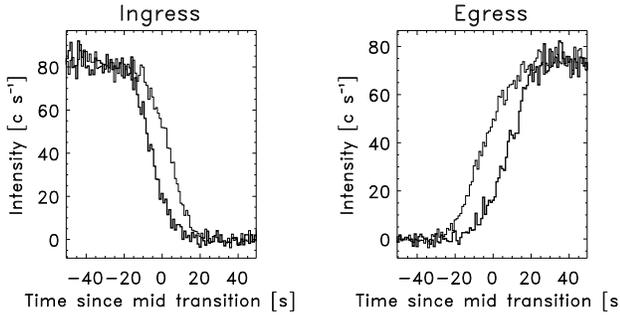,width=1.0\columnwidth,clip=t}
\caption{In thick lines, the average ingress profile (left) and egress
profile (right) in the 2--5~keV band; in thin lines in the 5--20~keV
band. The averages have been taken over 6 good-quality ingresses and 8
egresses. Background subtraction consisted of two steps: first for
particle-induced and cosmic background (using the {\tt pcabackest}
tool in {\tt ftools} version 5.2) and second for residual emission
during total eclipse.  The 5--20 keV intensities have been
renormalized to match the plateau values of the 2--5~keV curves.
\label{figfoldtrans}}
\end{figure}

In Fig.~\ref{figfoldtrans} we present the average ingress and egress
profiles for two separate bands: 2--5 keV and 5--20 keV. These show
that the eclipse lasts approximately 25~s longer in the low-energy
band, indicating absorption in the atmosphere of the obscuring
companion star. This is spectrally confirmed by the BeppoSAX
measurement (In~'t~Zand et al. 2000). 

We tested the transition profiles in the two bandpasses with a simple
model involving an isothermal spherical atmosphere in hydrostatic
equilibrium where the density is an exponential function of height,
assuming a constant source spectrum that is equal to that measured
with BeppoSAX (In 't Zand et al. 2000). This model turns out to give a
good representation of the data. Both the low and high-energy profiles
are well reproduced as is the lag between them. We find neutron star
traverse times over the scale height of $6.0\pm0.5$ and $8.5\pm0.5$~s
for the average ingress and egress respectively, with goodness-of-fits
of $\chi^2_{\rm red}=0.893$ (400 dof) and 0.989 (386 dof)
respectively. With the eclipse duration this results in upper limits
for the true scale height in terms of stellar radius of 0.2 to
0.3\%. We also tested linearly extended sources instead of a point
source, as would be expected for a nearly edge-on accretion disk, and
find fits that are only marginally better by about 3\% in $\chi^2_{\rm
red}$, with source-size equivalent traverse times of 14 and 20~s and
scale heights only 0.5 to 1.0~s smaller than for a point source.  We
regard the evidence for an extended source as weak.

\begin{figure}[t]
\psfig{figure=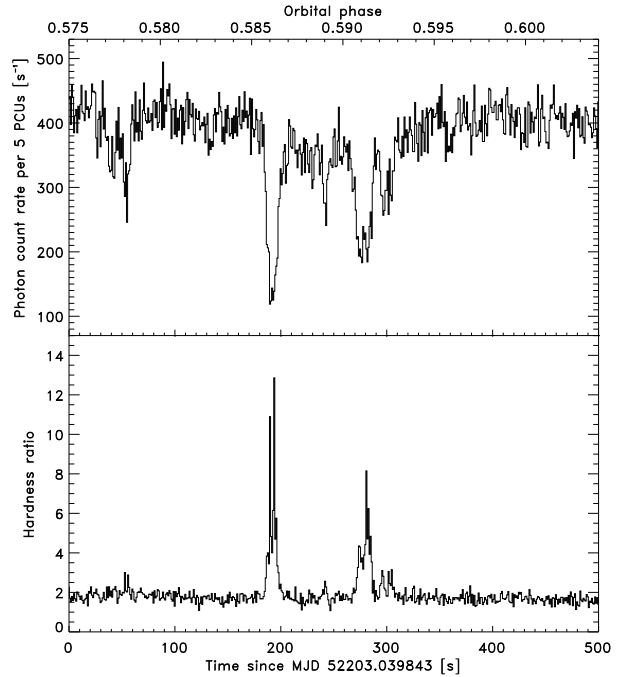,width=1.0\columnwidth,clip=t}
\caption{Upper panel: 2--20 keV light curve during a period of dipping
activity; lower panel: 5--10 to 2--5 keV hardness ratio. The times are
solar-system barycentered and the photon rates have been background
subtracted. The background count rate includes particle-induced
background, diffuse background and the local Galactic background as
determined from the full eclipse observed on MJD~52206 at which time
the same PCUs were employed.
\label{figdip}}
\end{figure}

As in other eclipsing low-mass X-ray binaries (LMXBs; EXO~0748-676,
MXB~1658-298, XTE~J1710-281), \bron\ shows dipping activity (for an
example, see Fig.~\ref{figdip}).  Dips are thought to be caused by
coverage of the emission region by irregular thickening of the
accretion disk (White \& Swank 1982). Thanks to the large inclination
angle, inferable from the occurrence of eclipses, the presence of dips
is not unexpected.  Dipping activity was observed during outbursts XI
(during two different binary orbits), XII (one orbit) and XIV (one
orbit). The orbital phase with respect to mid-eclipse was between 0.58
and 0.74.  The dip depth never exceeds 70\%, indicating that the
emission region is never completely obscured by the disk
irregularities. The dips last typically a fraction of a minute.

\section{Accurate positions of X-ray source and cluster center}
\label{pos}

{\rm Chandra\/} observed \bron\ on March 9, 2000, from 5:44:44 to
8:41:16 UT (PI J. Grindlay). This observation was triggered by our
RXTE observations showing the source to be active. The High Resolution
Camera HRC-I was used for a total net exposure time of 9964~s. No
grating was used.  The HRC-I is a microchannel plate detector with a
field of view of 30\arcmin$\times30$\arcmin\ and an angular resolution
of 0\farcs4 (FWHM), a bandpass of 0.1 to 10.0 keV with 227~cm$^2$
effective area at 1 keV, and minimal spectral resolution (Murray et
al. 2000). We extracted the data from the public archive for an
analysis of the position. Such an analysis was performed previously by
Revnitsev et al. (2002). However, we anticipated to improve on that by
attempting an astrometric solution of the image through X-ray
detections of cataloged stars.

\bron\ was significantly detected. A total of 15$\times10^3$ photons
were detected and the intensity was $1.50\pm0.01$~s$^{-1}$. Slow
variability was observed with a $\approx20$\% modulation around the
average on a time scale of 10$^3$~s.  Applying the 'celldetect' tool
from CIAO (version 2.2) to the image, we find four more point sources
with a net number of photons in excess of 10, and three sources
between 6 and 10. We determine the pixel positions of all eight
sources. Using the satellite aspect solution and requiring a match
within 1\arcsec\ (the absolute astrometric accuracy is quoted by the
Chandra team as 0\farcs6 to 0\farcs8 at 90\% confidence), we
tentatively identify three X-ray detections of stars in the USNO-B1.0
catalog (Monet et al. 2003). They are USNO0587-0575246 ($b_{\rm
USNO}=12.46, r_{\rm USNO}=11.45$), USNO0587-0575103 (14.79 and 13.32)
and USNO0586-0575381 (16.10 and 14.19). From the same optical image we
determined that the probability for chance coincidence of a star
brighter than $r_{\rm USNO}=14.19$ with one 1\arcsec-radius circle is
$7\times10^{-4}$ and we conclude that the identifications are
correct. We determined (X-ray minus optical) offset corrections of
$0\farcs68\pm0\farcs12$ and $0\farcs53\pm0\farcs12$ in
$\alpha_{2000.0}$ and $\delta_{2000.0}$ respectively. Applying these
corrections, we arrive at a position for \bron\ of
$\alpha_{2000.0}~=~17^{\rm h}50^{\rm m}46\farcs862$,
$\delta_{2000.0}~=~-31^{\rm o}16\arcmin28\farcs86$ with an error
radius of 0\farcs4 (95\% confidence). This position is consistent with
the position determined by Revnitsev et al. (2002), but has an 1.8
times better accuracy (when compared at the same confidence level).

\begin{figure}[t]
\psfig{figure=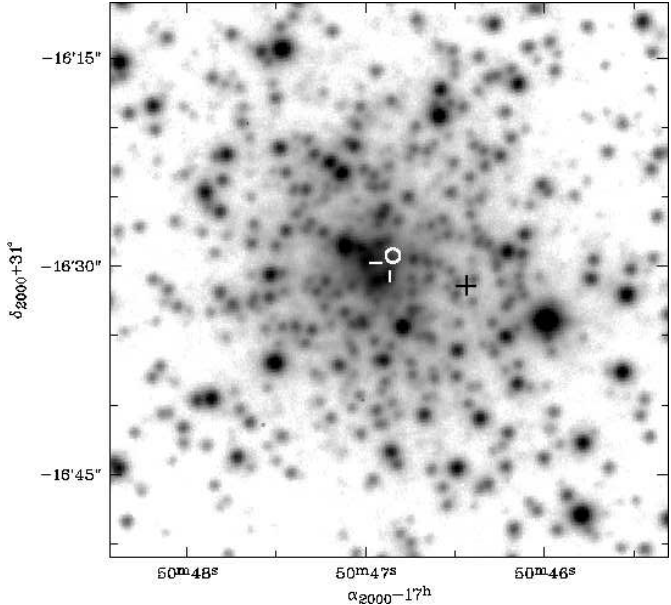,width=1.\columnwidth,clip=t,angle=270.}
\caption{Gunn~$z$ plate of Terzan 6 (Barbuy et al. 1997). The weakest
 stars are approximately $z=23$. The position of the bright transient
 is marked by a white circle (with error radius of 0\farcs5), the newly
 determined position of the center of gravity is indicated by two
 white lines, and the old position by a black cross.
\label{figgunnz}}
\end{figure}

On 17 May 1994, optical images of Terzan~6 were acquired with the
3.55~m New Technology Telescope (NTT) of the European Southern
Observatory, using the SUSI-1 camera which employed a 1024$\times$1024
CCD at the Nasmyth focus B, giving a plate scale of 0\farcs13 per
pixel (Barbuy et al.  1997). These images were taken under excellent
seeing conditions (0\farcs6) in $I$ (300~s), $V$ (480~s) and Gunn $z$
(120~s) filters. We carried out astrometry of the $V$ and $z$ plates
relative to the USNO-B1.0 catalog. Centroids of observed star profiles
were determined by Gaussian fitting. The astrometric solution
consisted of a zero point position, plate scale, and position angle on
the sky.  14 catalog stars were identified in the $V$ image with
deviations less than 0\farcs4, and 19 stars in the $z$ image with
deviations less than 0\farcs6. The rms values of the measured star
positions with respect to the cataloged positions is 0\farcs20 in
right ascension and 0\farcs14 in declination (0\farcs25 and 0\farcs24
respectively for $z$). Both solutions are less than 0\farcs15 apart.

The position of the center of Terzan~6 that is used throughout the
literature ($\alpha_{2000.0}~=~17^{\rm h}50^{\rm m}46\farcs44$,
$\delta_{2000.0}~=~-31^{\rm o}16\arcmin31\farcs4$) is due to the
original discovery plate (Terzan 1968, see also Terzan 1971). From
comparison with Fig.~\ref{figgunnz} it is clear that this position is
substantially off the densest concentration. Therefore, we
re-determined the center through the iterative centroiding method
introduced by Auriere (1982; see also Picard \& Johnston 1994). In
this method the (unweighted) average position is taken over all
resolved stars within a certain radius from a trial position. This
average becomes the trial position in the next iteration. The
procedure is repeated until the average converges. For Terzan~6, a
radius of 15\arcsec\ was chosen on the basis of the cluster's
appearance in our image. The resulting center is
$\alpha_{2000.0}~=~17^{\rm h}50^{\rm m}46\farcs854$,
$\delta_{2000.0}~=~-31^{\rm o}16\arcmin29.38\farcs4$ with an
uncertainty of 0\farcs6 as evaluated by testing different values for
the radius. The ultimate determination of the center is based on the
average position of 219 stars. This newly determined center is
5\farcs7 away from the old position. We note that this correction does
not affect the structural parameters of the cluster because those were
determined completely from relative positions (Trager et al. 1995).

The Chandra position of \bron\ is 0\farcs5 from our determination of
the center which is equivalent to 0.2 core radii. This confirms the
suggestion from the sub-day orbital period and the moderately high
collision number (Verbunt 2001) that this binary was formed due to
tidal capture or exchange encounter.

No sources belonging to Terzan~6 other than \bron\ were
unambiguously detected in the Chandra image.  Outside the point spread
function of \bron, the upper limit for the intrinsic luminosity for a
0.3~keV black body spectrum with $N_{\rm
H}=1.4\times10^{22}$~cm$^{-2}$ is approximately
$1.3\times10^{33}$~\lum. The point-spread function is roughly given by
a Gaussian profile with a standard deviation of 0\farcs4. It degrades the
sensitivity to faint sources by a factor of more than two for
distances closer to \bron\ than 1\farcs8.

\section{Discussion}
\label{discuss}

\subsection{Outburst behavior}

The 2--10 keV peak flux is mostly about 7$\times10^{-10}$~\ecs.
Broad-band measurements during outburst VIII allowed an extrapolation
to 0.1--200 keV of 1.0$\times10^{-9}$~\ecs\ (In 't Zand et
al. 2000). The outburst duration is always 4~weeks. The
fractional rms in the average intensity from outburst to outburst is
5\%. If we assume that the spectrum does not change drastically
from outburst to outburst, the luminosity averaged over
all times (outburst as well as quiescence) is 1.5$\times10^{36}$~\lum\
and the equivalent mass accretion rate, assuming 100\% efficiency of
transforming the liberated gravitational energy to radiation, is about
1.2$\times10^{-10}$~M$_\odot$yr$^{-1}$ (i.e., 1\% of Eddington). One
accurately measured outburst (XII) was delayed by 25\% of the average
wait time. The peak of that outburst was also 25\% larger while the
duration was the same, so that the average luminosity remained
constant. It must be a rather constant mass transfer rate from the
companion star to the accretion disk which is responsible for the
regular outburst behavior. A similar behavior is observed in the Rapid
Burster in the GC Liller~1: it shows a quasi-periodic outbursts whose
period of about 210 days suddenly switched to 100 days in Nov. 1999 in
combination with a factor-of-two decrease in the peak flux (Masetti
2002). Another well-documented case is Aql~X-1, with at least 31
outbursts detected thus far (e.g., \u{S}imon 2002), but its recurrence
time is not so well defined and there does not seem to be as strong a
correspondence between recurrence time and fluence per outburst.

The transient nature that many LMXBs exhibit is generally thought to
be related to thermal-viscous instabilities in the accretion disk
where the thermal conditions are predominantly determined by heating
through X-rays from the central object (e.g., Smak 1983; Osaki 1996;
van Paradijs 1996; Lasota 2001, \u{S}imon 2002). These instabilities
become important and the source becomes transient if the average X-ray
luminosity is below a critical value (van Paradijs 1996, see also
White et al. 1984). The onset of an outburst occurs if the surface
column density at any location in the accretion disk exceeds a
critical value. As a consequence the temperature rises, the viscosity
increases, and heating fronts travel in and outward through the
disk. Depending on the radius of the ignition, differently sized
outbursts may result (outward fronts are expected to be less
efficient, possibly resulting in smaller outbursts if the igition
occurs in the inner parts of the disk).  Furthermore, the ignited
parts of the disk may irradiate other cool parts of the disk and
ignite those. The increased viscosity results in a depletion of the
accretion disk which may be complete or partial. The latter is more
likely to happen if the ignition radius is small. The recurrence time,
peak flux and outburst duration depend on quite a number of parameters
and effects, but it seems obvious that the quasi-periodicity at least
testifies to a rather stable mass transfer rate from the
secondary. The similar recurrence times of \bron, Aql~X-1 and the
Rapid Burster probably reflect similar orbital periods (19~hrs for Aql
X-1, unknown for the Rapid Burster), similar mass transfer rates and
the fact that they are all neutron star systems without considerable
truncation of the disks that may result in decades-long recurrence
times such as in black hole systems (e.g., Dubus et al. 2001).

\subsection{Eclipse timing and evolutionary constraints}

As illustrated by Horne et al.\ (1985), the eclipse duration expressed
as a fraction of the orbital period constrains the mass ratio and
inclination. For the low-mass X-ray binary in Terzan 6 we find that
the inclination must be higher than $74\fdg5$, and the mass of the mass
donor must be higher than 0.07 times the mass of the neutron
star. Thus, for a canonical 1.4~M$_\odot$ neutron star, the lower
limit on the mass of the secondary is 0.1~M$_\odot$; a likely upper
limit is 0.8~M$_\odot$.  Kepler's third law yields a semi-major axis
of 3.0 to 3.6~R$_\odot$ for these masses. This implies an average
orbital velocity of the secondary with respect to the primary between
296 and 356 km~s$^{-1}$. 

%Since the secondary must be filling its Roche
%lobe in order to invoke mass transfer, its radius must be between 0.7
%(for a 0.1~M$_\odot$ secondary) and 1.2~R$_\odot$ (for 0.8~M$_\odot$;
%following Eggleton 1983). This is between 1.4 and 5.5 times the main
%sequence radius (this radius does not differ significantly for
%population I and II stars; e.g., Tout et al. 1996). For certain, the
%secondary is not a helium main sequence star or white dwarf. The Roche
%lobe size would match the main sequence radius if the secondary has a
%mass of 1.4~M$_\odot$. This mass would be smaller if the neutron star
%is lighter, but that is contrary to what is expected.

The eclipse timing resolves an orbital period that has remained
constant within $|\dot{P}/P|<3\times10^{-8}$~yr$^{-1}$ for almost 4
years.  This corresponds to a lower limit to the time scale of any
orbital period change of $3\times10^7$~yr which is mildly
constraining. Only one LMXB has evidence of a shorter time scale:
$P/\dot{P}=2\times10^7$ yr in EXO~0748-676 (Wolff et al. 2002).

%Orbital period evolution is expected due to
%gravitational radiation of the two orbiting bodies and magnetic
%braking (Bhattacharya \& van den Heuvel 1991; Verbunt 1993).  For
%conservative mass transfer in systems where the secondary star is
%non-degenerate, $\dot{P}/P = \dot{M}_2/M_2$, where $M_2$ is the
%mass of the secondary star.  Thus, based on these assumptions, the
%orbital period is expected to {\it decrease\/} on time scales of
%$|P/\dot{P}| \sim 10^{8}-10^{10}$ yr.  However, it has been noted
%that in a number of systems the time scale is considerably shorter
%than this, and in some cases the orbital period is seen to be {\it
%increasing}.

We may compare these observational constraints with theory as follows.
In a globular cluster, the most massive star still on the main
sequence has a mass of about 0.8~M$_\odot$.  A main-sequence star of
this mass, or lower, cannot fill its Roche lobe in an orbit of
0.515\,d.  Thus, the donor of \bron\ must be larger than a
main-sequence star of the same mass.  If it is larger due to stellar
evolution into a (sub)giant, then its initial mass was the turnoff
mass of 0.8~M$_\odot$.  For a neutron-star mass of 1.4~M$_\odot$ and a
companion of 0.8~M$_\odot$, the orbital period gives a distance
between the two stars of 3.6 R$_\odot$, and a Roche lobe for the donor
of 1.2 R$_\odot$.  At such a radius, a subgiant expands on a time
scale longer than $R/\dot R\simeq10^{10}$\,yr.  If the donor has
already reduced its mass by mass transfer to the neutron star, it
Roche lobe will be somewhat smaller (e.g.\ 0.9 R$_\odot$ at a mass of
0.4~M$_\odot$), and the expansion time scale longer.  For conservative
mass transfer, in which no mass or angular momentum is lost from the
binary, the time scales for the mass transfer and for changes in the
orbital period are set by the expansion time scale of the donor, i.e.\
$M/\dot M\sim P/\dot P\sim R/\dot R$. The time averaged mass transfer
rate of $\dot M\simeq 1.4\times10^{-10} M_{\odot}$~yr$^{-1}$ derived above,
and the limit of the period change $\dot P/P<3\times10^{-8}$~yr$^{-1}$ are
both compatible with the expansion rate of a subgiant donor star.  The
low mass transfer rate furthermore indicates that the time scale for
angular momentum loss from the binary through magnetic braking is
longer than $\sim 10^{10}$\,yr.

An alternative explanation for an expanded donor star would be that
the donor still contains a fair amount of thermal energy left over
from its capture by the neutron star (e.g., Verbunt 1994). Since the
life time of such a source would be set by its thermal time scale,
this possibility is less probable a priori; and the arguments above
show that it is not necessary as all the properties of the low-mass
X-ray binary may be explained with an ordinary subgiant donor.

\subsection{Eclipse profile}

The transition profiles in \bron\ are remarkably stable, when compared
to other LMXBs. Ingress and egress durations range from -30 to +30\%
with respect to the average. In EXO~0748-676, for instance, the
variation is 100\% (Wolff et al. 2002). The stability may be due to
either a relative stability of magnetic activity on the secondary, or
a relative smoothness of the accretion process (and associated X-ray
irradiation of the secondary).

We observed a difference in the eclipse duration at 2--5 keV and
5--20~keV energies and modeled this by assuming an isothermal
spherical atmosphere in hydrostatic equilibrium where the density is
an exponential function of height. This simple model reproduces the
eclipse duration difference for a scale height, as projected on the
path of the neutron star line of sight through the atmosphere, of
$2.0\times10^3$~km for the ingress and $2.8\times10^3$~km for the
egress. The model reproduces the curvature on both ends of the
transition profiles and the spectral change over the transition. The
true scale height depends on the precise inclination angle. If the
primary and secondary are 1.4 and 0.8~M$_\odot$ respectively, the
inclination angle would be 74\fdg5, the secondary radius 1.2~R$_\odot$
and the true scale height $1.2\times10^3$~km. This is about 2 times
larger than expected, but this could be explained by X-ray irradiation
of the secondary.

The $\approx25$\% difference between the average ingress and egress
duration has not been reported before in any other LMXB
eclipser. Perhaps the relative stability of the transitions is what
makes this measurement possible in this particular source. The
difference is indicative of a similar difference in the atmosphere's
scale height between the leading and trailing hemisphere of the
secondary.  Day et al. (1988) predicted such a scale height
difference, in the same sense, from a Coriolis force acting on
supersonic flows on the secondary's surface as invoked by X-ray
irradiation. This will increase the scale height on the trailing
hemisphere while decreasing it on the leading hemisphere. Whether this
model is viable to \bron\ is hard to verify without the detection of
the secondary star (note that the secondaries have been detected in
the well-documented eclipsers Her X-1, MXB~1658-298 and EXO 0748-676).

\subsection{Bursts}

The first-time detection of type-I X-ray bursts unambiguously proves
that the compact object in \bron\ is a neutron star. This leaves AC211
in M15 as the sole bright LMXB in a Galactic globular cluster for
which no nature has been determined of the compact object. Finding
bursts in this object is going to be at least as difficult as for
\bron, because it also is a high-inclination system. In fact, the
central source is obscured by the accretion disk rim {\em all the
time} (e.g., Ilovaisky et al. 1993) and bursts can only be detected
indirectly through scattered X-rays.  Another complication for AC211
is that it is accompanied by another bright and bursting LMXB at a
distance of only 2\farcs7 (White \& Angelini 2001) so that only
Chandra and XMM-Newton will be able to localize bursts accurately
enough.

All seven bursts from \bron\ are relatively short suggesting that the
flashes occur in a hydrogen-poor/helium-rich layer which, according to
burst theory (Fujimoto et al. 1981), must have been formed by stable
hydrogen fusion. For that to happen, the accretion rate must be higher
than about 1\% of Eddington. This value is consistent with the
observations.  The highest observed peak flux translates to a
luminosity of $(2.3^{+1.9}_{-1.1})\times10^{38}$~\lum\ for a distance
of 9.5$^{+3.3}_{-2.5}$~kpc (Kuulkers et al. 2003). This value is
consistent with the Eddington limit as expected for a hydrogen-rich
atmosphere. The four bursts that were well measured with the PCA have
the following interesting characteristics:
\begin{enumerate}
\item
there is a large range of peak fluxes from 0.2 to
$2.1\times10^{-8}$~\ecs;
\item
the two radius expansion bursts have fast rise times below 1~s, while
the others have slower rise times of at least 2~s. In fact, 
burst no. 1 has a rise and decay of similar duration which is
atypical for a type-I burst;
\item
one radius-expansion burst (no. 3) has a 1.8 times higher
unabsorbed bolometric peak flux than the other;
\item
the minimum apparent radius of a spherical blackbody is as low at 4~km.
\end{enumerate}
We suspect that at least three of these atypical observations are
associated with the high inclination angle of the binary. This
suspicion is motivated by a recent observation of MXB~1658-298 by
Wijnands et al. (2002). They observed two bursts during a period of
dipping activity which had peak intensities that were four and ten
times smaller than of most ordinary bursts and which showed slow rises
that were similar to the decay times. We should point out that there
is one characteristic which makes the two MXB~1658-298 bursts
different from the \bron\ ones: they were separated by merely (a
record) 50~s which suggests that incomplete burning (e.g., Fujimoto et
al. 1987) may have been important in determining some of the burst
characteristics.  Short recurrence times can be excluded for \bron:
the data coverage around the two small bursts without additional
bursts is at least 1200~s. However, it is our impression that the
short recurrence time does not explain the slow rise times. No dipping
was observed around the bursts of \bron, but we note that the two
small bursts were observed at orbital phases of 0.87 (burst 1) and
0.49 (burst 4) that are more favorable for obscuration effects than
for the bright bursts (the phase is 0.28 for burst 2 and 0.21 for
burst 3).

Regarding the factor-of-1.8 peak flux difference between the two
radius expansion bursts, this is similar as in another well-measured
system. Galloway et al. (2003) were able to gather observations of 61
photospheric radius expansion bursts from 4U~1728-34 (=GX~354-0). They
found a variation in the peak flux by a factor of 1.5 which is
correlated with spectral variations of the persistent emission. They
suggest that a variation in the geometry of the accretion disk may be
at work, for instance precession of a warped disk.

\acknowledgement We are indebted to Evan Smith for his tireless
efforts to schedule the many time-constrained TOO observations that
led to this paper. JZ and EK acknowledge financial support from the
Netherlands Organization for Scientific Research (NWO).


\begin{thebibliography}{99}

\bibitem{angel}Angelini, L., Loewenstein, M., \& Mushotzky, R.F. 2001, ApJ, 557, L35
\bibitem{aur} Auriere, M. 1982, A\&A, 109, 301
\bibitem{barbuy}Barbuy, B., Ortolani, S., \& Bica, E. 1997, A\&A, 122, 483
\bibitem{boe97a} Boella, G., Butler, R.C., Perola, G.C., Piro, L.,
      Scarsi, L., \& Bleeker, J.A.M., 1997, A\&AS, 122, 299
\bibitem{coc} Cocchi, M., Bazzano, A., Natalucci, L., et al. 2001,
       A\&A, 387, L37
\bibitem{comi} Cominsky, L.R. \& Wood, K.S. 1984, ApJ, 283, 765
\bibitem{day}Day, C.S.R., Tennant, A.F., \& Fabian, A.C. 1988, MNRAS, 231, 69
\bibitem{dub01} Dubus, G., Hameury, J.-M., \& Lasota, J.-P. 2001, A\&A, 373, 251
% \bibitem{eggl} Eggleton, P.P. 1983, ApJ, 268, 368
\bibitem{fuk} Fujimoto, M.Y., Hanawa, T., \& Miyaji, S. 1981, ApJ, 247, 267
\bibitem{fuk2} Fujimoto, M.Y., Sztajno, M., Lewin, W.H.G., \& van Paradijs, J.
        1987, ApJ, 319, 902
\bibitem{gall} Galloway, D.K., Psaltis, D., Chakrabarty, D., \& Muno, M.P.
        2003, ApJ, submitted (astro-ph/0208464)
%\bibitem{gott}Gottwald, M., Haberl, F., Parmar, A.N., \& White, N.E.
%        1986, ApJ, 308, 213
\bibitem{horne} Horne, K. 1985, MNRAS, 213, 129
\bibitem{ilo} Ilovaisky, S.A., Auxi\`{e}re, M., Koch-Miramond, L., Chevalier, C.,
        Cordoni, J.-P., \& Crow, R.A. 1993, A\&A, 270, 139
\bibitem{intzand1}In 't Zand, J.J.M. 2001, In Proc. 4th INTEGRAL workshop
        (Sep. 2000, Alicante, Spain), eds. A, Giminez, V. Reglero \& C.
       Winkler, 463
\bibitem{zand1}In 't Zand, J.J.M., Kuulkers, E., Bazzano, A., et al. 2000,
        A\&A, 355, 145
%\bibitem{zand2}In 't Zand, J.J.M., Cornelisse, R., Kuulkers, E. 2001,
%        A\&A, 372, 916
%\bibitem{zand3}In 't Zand, J.J.M., Verbunt, F., Kuulkers, E., et al. 2002,
%        A\&A, 389, L43
\bibitem{Jag}Jager R., Mels, W.A., Brinkman, A.C., et al., 1997, A\&AS, 125, 557
\bibitem{jaho}Jahoda, K., Swank, J.H., Stark, M.J., et al. 1996, Proc. SPIE,
        2808, 59
\bibitem{kennea} Kennea, J.A., \& Skinner, G.K. 1996, PASJ, 48, L117
\bibitem{kundu}Kundu, A., Maccarone, Th \& Zepf, S. 2002, ApJ, 574, L5
\bibitem{kuulk} Kuulkers, E., Hartog, P.R., In 't Zand, J.J.M., et al. 2003,
        A\&A, 399, 663
\bibitem{lasota} Lasota, J.-P. 2001, NewAR, 45, 449
\bibitem{lewin}Lewin, W.H.G., van Paradijs, J., \& Taam, R.E. 1993, Space Sci.
        Rev., 62, 223
\bibitem{levine} Levine, A.M., Bradt, H., Cui, W., et al. 1996, ApJ, 469, L33
\bibitem{maeda}Maeda, Y., Koyama, K., Sakano, M., Takeshima, T., \& Yamauchi,
        S. 1996, PASJ, 48, 417
\bibitem{mara} Maraschi, L., \& Cavaliere, A. 1977, in Highlights in
        Astron., 4, 127
\bibitem{markw0}Markward, C.B., Swank, J. \& Marshall, F.E. 1999, IAUC 7120
%\bibitem{markw2}Markwardt, C.B., Swank, J., Strohmayer, T.E., In 't Zand,
%        J.J.M. \& Marshall, F.E. 2002, ApJ, 575, L21
\bibitem{markw1}Markward, C.B., \& Swank, J. 2002, presented at the
        April 2002 meeting of the High-Energy Astrophysics Division meeting
        in Albuquerque, abstract X11.007
\bibitem{maset}Masetti, N. 2002, A\&A, 381, L45
\bibitem{monet}Monet, D., Levine, S., Canzian, B., et al. 2003, ApJ,
        in press (astro-ph/0210694)
\bibitem{murr} Murray, S.S., Austin, G., Chappell, J., et al. 2000,
        Proc. SPIE, 4012, 68
\bibitem{oost}Oosterbroek, Parmar, A.N., Sidoli, L., In 't Zand, J.J.M.,
        \& Heise, J. 2001, A\&A, 376, 532
\bibitem{osaki} Osaki, Y. 1996, PASP, 108, 39
\bibitem{parm}Parmar, A.N., White, N.E., Giommi, P. \& Gottwald, M. 1986,
        ApJ, 308, 199
\bibitem{pavl}Pavlinsky, M.N., Grebenev, S.A., \& Sunyaev, R.A. 1994, ApJ, 425, 110
\bibitem{pic} Picard, A., \& Johnston, H.M. 1994, A\&A, 283, 76
\bibitem{pred} Predehl, P., Hasinger, G., \& Verbunt, F. 1991, A\&A, 246, L21
\bibitem{rev}Revnitsev, M.G., Trudolyubov, S.P., \& Borozdin, K.N. 2002,
        Astron. Lett., 28, 276
\bibitem{sara}Sarazin, C., Irwin, J., \& Bregman, J. 2001, ApJ, 556, 533
\bibitem{simon} \u{S}imon, V. 2002, A\&A, 151, 167
\bibitem{smak} Smak, J. 1983, ApJ, 272, 234
\bibitem{stroh1}Strohmayer, T.E., Jahoda, K., Giles, A.B., \& Lee, U. 1997,
        ApJ, 486, 355
\bibitem{stroh2}Strohmayer, T.E., \& Bildsten, L. 2003, in ``Compact Stellar
        X-Ray Sources, eds. W.H.G. Lewin and M. van der Klis, Cambridge
        University Press, in press (astro-ph/0301544)
\bibitem{Swank2}Swank, J., \& Markwardt, C. 2001, In Proc. ``New Century
        of X-ray Astronomy'', eds. H. Inoue \& H. Kunieda, PASP Conf. Ser.,
        p. 94
\bibitem{Tanan}Tananbaum, H., et al. 1972, ApJ, 174, L143
\bibitem{terz1} Terzan, A. 1968, C.r. Acad. Sci. Paris 267, s\'{e}rie B, 1245
\bibitem{terz2} Terzan, A. 1971, A\&A, 12, 477
%\bibitem{Tout} Tout, C.A., Pols, O.R., Eggleton, P.P., \& Han, Z. 1996,
%        MNRAS, 281, 257
\bibitem{trag} Trager, S.C., King, I.R., \& Djorgovski, S. 1995, AJ, 109, 218
\bibitem{jvp2} Van Paradijs, J.,  1996, ApJ, 464, L139
\bibitem{vaugh} Vaughan, B.A., van der Klis, M., Wood, K.S., et al.
        1994, ApJ, 435, 362
\bibitem{verb0} Verbunt, F. 1994, A\&A, 285, L21
\bibitem{Verb} Verbunt, F. 2001, in Proc. '$\omega$ Cen, a unique window
        in astrophysics', eds. F. van Leeuwen, G. Piotto and
        J. Hughes, ASP Conf. Ser, p. 289 (astro-ph/0111441)
%\bibitem{Verb1} Verbunt, F. 2002, in Proc. 'New horizons in globular cluster
%        astronomy', eds. G. Piotto, G. Meylan, G. Djorgovski, M. Riello,
%        ASP Conf. Ser, in press (astro-ph/0210057)
\bibitem{verb2} Verbunt, F., Bunk, W., Hasinger, G., \& Johnston, H.M. 1995,
        A\&A, 300, 732
\bibitem{whi1}White, N. E., \& Swank, J. H. 1982, ApJ, 253, L61
\bibitem{whi3}White, N. E., Swank, J. H., \& Kaluzienski, J.L. 1984,
        in High Energy Transients in Astrophysics, ed. S. Woosley, AIP, p. 31
\bibitem{whi2}White, N.E. \& Angelini, L. 2001, ApJ, 561, L101
\bibitem{wijn}Wijnands, R., Muno, M.P., Miller, J.M., Franco, L.M.,
        Strohmayer, T., Galloway, D., \& Chakrabarty, D. 2002, ApJ, 566, 1060
\bibitem{wolff}Wolff, M.T., Hertz, P., Wood, K.S., Ray, P.S., \& Bandyopadhyay,
        R. M. 2002, ApJ, 575 384
\bibitem{woos} Woosley, S.E., \& Taam, R.E. 1976, Nat, 263, 101
\end{thebibliography}
\end{document}